\documentclass[%
 reprint,
 amsmath,amssymb,
 aps,
]{revtex4-1}

\usepackage{graphicx}% Include figure files
\usepackage{dcolumn}% Align table columns on decimal point
\usepackage{bm}% bold math

\usepackage{color}
\newcommand{\GeV}{\makebox{ GeV}}
\newcommand{\beq}{\begin{equation}}
\newcommand{\enq}{\end{equation}}
\newcommand{\beqa}{\begin{eqnarray}}
\newcommand{\beqast}{\begin{eqnarray*}}
\newcommand{\enqa}{\end{eqnarray}}
\newcommand{\enqast}{\end{eqnarray*}}

\def\GeV{\nobreak\,\mbox{GeV}}

\begin{document}

%\preprint{APS/123-QED}

\title{Energy Dependence  of Profile Functions in $\rm{p\bar p}$ and $\rm pp$ Scattering }
 
 \author{A. K. Kohara} 
 \affiliation{Instituto de F\'{\i}sica, Universidade Federal do Rio de
 Janeiro \\
 C.P. 68528, Rio de Janeiro 21945-970, RJ, Brazil   }
\author{E. Ferreira} 
 \affiliation{Instituto de F\'{\i}sica, Universidade Federal do Rio de
 Janeiro \\
 C.P. 68528, Rio de Janeiro 21945-970, RJ, Brazil   }
\author{T. Kodama} 
\affiliation{Instituto de F\'{\i}sica, Universidade Federal do Rio de
 Janeiro \\
 C.P. 68528, Rio de Janeiro 21945-970, RJ, Brazil   }

\date{\today}% It is always \today, today,

\begin{abstract}

We construct analytical forms in the impact parameter $b$-space for the real 
and imaginary amplitudes describing elastic pp and $\rm p \bar p $ scattering. 
The amplitudes converted analytically to the momentum transfer $t$-space have 
magnitudes, slopes, curvatures, zeros, signs, obeying phenomenological 
and theoretical expectations, and describe with high precision all 
details of the data, in the full-t range, and for energies from 20 GeV to 
7 TeV.  The connection of forward with large-$|t|$ behavior allows
precise determination of total cross-sections, slopes and other 
scattering parameters. We study the properties and the energy dependence 
of the $b$-space profile functions, observing that the real part has 
fundamental influence in the structure of $d\sigma/dt$ at intermediate 
and large $|t|$ values.  We discuss the 540/546 GeV and 1.8/1.96 TeV data 
from CERN SPS and Fermilab TEVATRON and the 7  TeV  results from TOTEM 
measurements at LHC, and investigate the extrapolation to 14 TeV and 
higher energies. 

\end{abstract}

\maketitle

%\tableofcontents

\section{Introduction and motivation \label{sec-introduction}}

The differential cross sections of elastic pp and $\mathrm{{p\bar{p}}}$
scattering are described by two amplitudes, $T_{R}(s,t)$ and $T_{I}(s,t)$,
through 
\begin{equation}
\frac{d\sigma }{dt}=\left( \hbar c\right) ^{2}~|T_{R}(s,t)+iT_{I}(s,t)|^{2}~.
\label{dsigdt}
\end{equation}
Two amplitude functions contain basic information on the proton
structure and are represented by two kinematical variables $(s,t)$ only: 
this simplicity gives charm and importance of the elastic scattering 
process. In spite of the this simplicity, and of the 40 years of 
measurements and theories, the real and imaginary terms are not known 
well enough.
Microscopic models have not been precise and clarifying enough for the
separate identification of the two parts. In the present work, the
data on pp and p$\mathrm{\bar{p}}$ elastic scattering at high energies 
are described in terms of real and imaginary amplitudes 
with high accuracy  in the  whole t-range, within a
formalism including  the expectation of dispersion relation for amplitudes
and for slopes. This treatment leads also to a determination of the total
cross section and other forward scattering parameters, permitting to
investigate the compatibility of data and estimates. 
The identification of real and imaginary parts leads to values for 
slopes  $B_R$ and $ B_I$ of the real and imaginary amplitudes, 
that are independent quantities, influence the amplitudes in the whole 
t-range and are important for the determination of the total cross
section. The amplitudes are fully constructed analytically both in 
impact parameter and momentum transfer spaces, and investigation is 
made of their extension to very large $|t|$, to include the universal
contribution of the perturbative process of three gluon exchange.

The analysis shows the importance of the real part that, although having
small contribution to the  integrated elastic cross section, is crucial 
in the description of the details of the data at intermediate and
high $|t|$.  Our description leads to a prediction for a marked dip
in $d\sigma /dt$ of p$\mathrm{\bar{p}}$ elastic scattering in the $|t|$
range 3 - 5 GeV$^{2}$ at the energies 0.546 and 1.8 TeV due to the
cancellation between the perturbative real term of negative sign and the
real non-perturbative contribution of positive sign in this region.

 Coulomb interference is treated properly, with the Coulomb phase 
derived taking into account the difference in slopes of the 
real and imaginary amplitudes. In the 541/546 GeV case, the data of
event rate at very low $|t|$  are submitted to the Coulomb interference 
forms, and normalization connection with absolute differential cross 
section is obtained. The low $|t|$ data for dsigma/dt thus constructed 
is joined to the data of other measurements, and a continuous and 
consistent data-basis is obtained for analysis. 

In the 1.8/1.96 TeV case, the forward scattering data of the E-710 and 
CDF experiments are joined to the recent D0 Collaboration points at 
large $|t|$, and attempts are made to find unified treatments of the 
experimental information.

\section { \label{basic_forms} Analytical form of profile functions }
%   in  $\rm pp$  and $\rm p\bar p$ scattering } 

To establish a bridge connecting data to theory, the amplitudes $T_{R}$ 
and $T_{I}$ must be determined phenomenologically as functions of 
$s$ and $t$. This is a disentanglement problem, which in principle 
has no completely 
model-independent solution. However, we believe that our efforts lead to,
possibly realistic, representations for the amplitudes, as functions of $|t|$
in their full intervals, for each of the measured energies with minimal
dependence on a specific model.

In practical terms, we need to propose analytical representations for imaginary and real parts,
proving that they cover with accuracy all $|t|$ ranges present in the data.
At a given energy, the determination of fundamental parameters such as total
cross section, slopes, $\rho$ ratio, depend on limits $|t| \rightarrow 0 $
evaluated with specific functions $T_R(s,t)$ and $T_I(s,t)$. These forms
must have curvatures, zeros, magnitudes, signs, that build the whole
observed $d\sigma/dt$ structure. It is fundamental that precise reproduction
of the data be obtained, with regular evolution between neighbor
experimental energies. Theoretical constructions should be checked against
the imaginary and real terms, which must be considered as necessary and
reliable bridge between data and models.

We propose   \cite{ferreira1}    analytical forms in the 
impact parameter $b$-space   
\begin{equation}
\label{form_b_1}
 \tilde{T}_{K}(s,b)=\frac{\alpha_{K}}{2~\beta_{K}}e^{-\frac{b^{2}}{4\beta_{K}}%
}+\lambda_{K}~\tilde{\psi}_{K}(s,b)~,
 \end{equation}
where
\begin{equation}
\label{shape}
\tilde{\psi}_{K}(s,b)=\frac{2~e^{\gamma_{K}}}{a_{0}}~\frac{e^{-\sqrt
{\gamma_{K}^{2}+\frac{b^{2}}{a_{0}}}}}{\sqrt{\gamma_{K}^{2}+\frac{b^{2}}%
{a_{0}}}}\Big[1-e^{\gamma_{K}}~e^{-\sqrt{\gamma_{K}^{2}+\frac{b^{2}}{a_{0}}}%
}\Big]~
\end{equation} 
is a shape function. The indices $K=I$ and $K=R$ in $\tilde{T}_{K}(s,b)$ 
refer to the real and imaginary parts. The form introduced in 
Eq.(\ref{shape}) has origin in studies with  the Stochastic Vacuum Model 
\cite{dosch},  and four energy-dependent parameters have been introduced 
for each amplitude.

We note that  $\tilde{\psi}_{K}(s,b=0) ~ = ~ 0 $, so that    
 $ \tilde{T}_{K}(s,b=0)~ =~ {\alpha_{K}}/{2~\beta_{K} }   $ .

The amplitudes $T_R(s,t)$ and $T_I(s,t)$  given as functions of the
momentum transfer $|t|$ are obtained through Fourier transforms
\begin{equation}
\tilde{T}_{K}(s,b)=\frac{1}{2\pi}\int d^{2}\vec{q}~e^{-i\vec{q}.\vec{b}}%
~T_{K}^{N}(s,t=-q^{2})~.
\end{equation}

One of the advantages of this
representation is that it permits  analytic representation also in 
$t$ space  \cite{ferreira1}.  

  The connection with the quantities used in the  description of forward 
scattering  are 
\begin{equation}
\sigma(s)=4\sqrt{\pi}\left(  \hbar c\right)  ^{2}~(\alpha_{I}(s)+\lambda
_{I}(s))~,\label{sigma_par}%
\end{equation}%
\begin{equation}
\rho(s)=\frac{T_{R}^{N}(s,t=0)}{T_{I}^{N}(s,t=0)}=\frac{\alpha_{R}%
(s)+\lambda_{R}(s)}{\alpha_{I}(s)+\lambda_{I}(s)}~,\label{rho_par}%
\end{equation}
and 
\begin{align}
B_{K}(s) =\frac{1}{T_{K}^{N}(s,t)}\frac{dT_{K}^{N}(s,t)}{dt}\Big|_{t=0}%
=~\frac{1}{\alpha_{K}(s)+\lambda_{K}
s)}\times\nonumber\\
\Big[\alpha_{K}(s)\beta_{K}(s)+\frac{1}{8}\lambda_{K}(s)a_{0}\Big(6\gamma
_{K}(s)+7\Big)\Big]~ .\label{slopes_par}%
\end{align}

We stress the importance of different values allowed for the imaginary 
and real parts of the complex amplitude, as equired by dispersion 
relations \cite{ferreira2}.
 
In the numerical  analysis of data we  work  directly with the amplitudes 
in $t$-space because they are more directly connected to the  
$d\sigma/dt$ measurements. 
Using as units milibarns for cross sections and $\GeV^2$ for the 
momentum transfer squared $t$, we have 
$ \left(  \hbar c\right) ^{2}~=~ 0.389 ~ {\rm mb} ~\GeV^2$.

These expressions  represent the non-perturbative dynamics  of 
scattering for all $|t|$ . For very large $|t|$ there is also a 
contribution of a perturbative tail representing 3-gluon exchange, which 
is assumed to be universal, independent of the energy. It  has been first 
observed \cite{Faissler} at 27.4 GeV, and gives a $|t|$ dependence of 
the form $1/|t|^8$ in $d\sigma/dt$ \cite{{DL_3g}}.  

\begin{figure}[b]
\includegraphics[width=8.3cm]{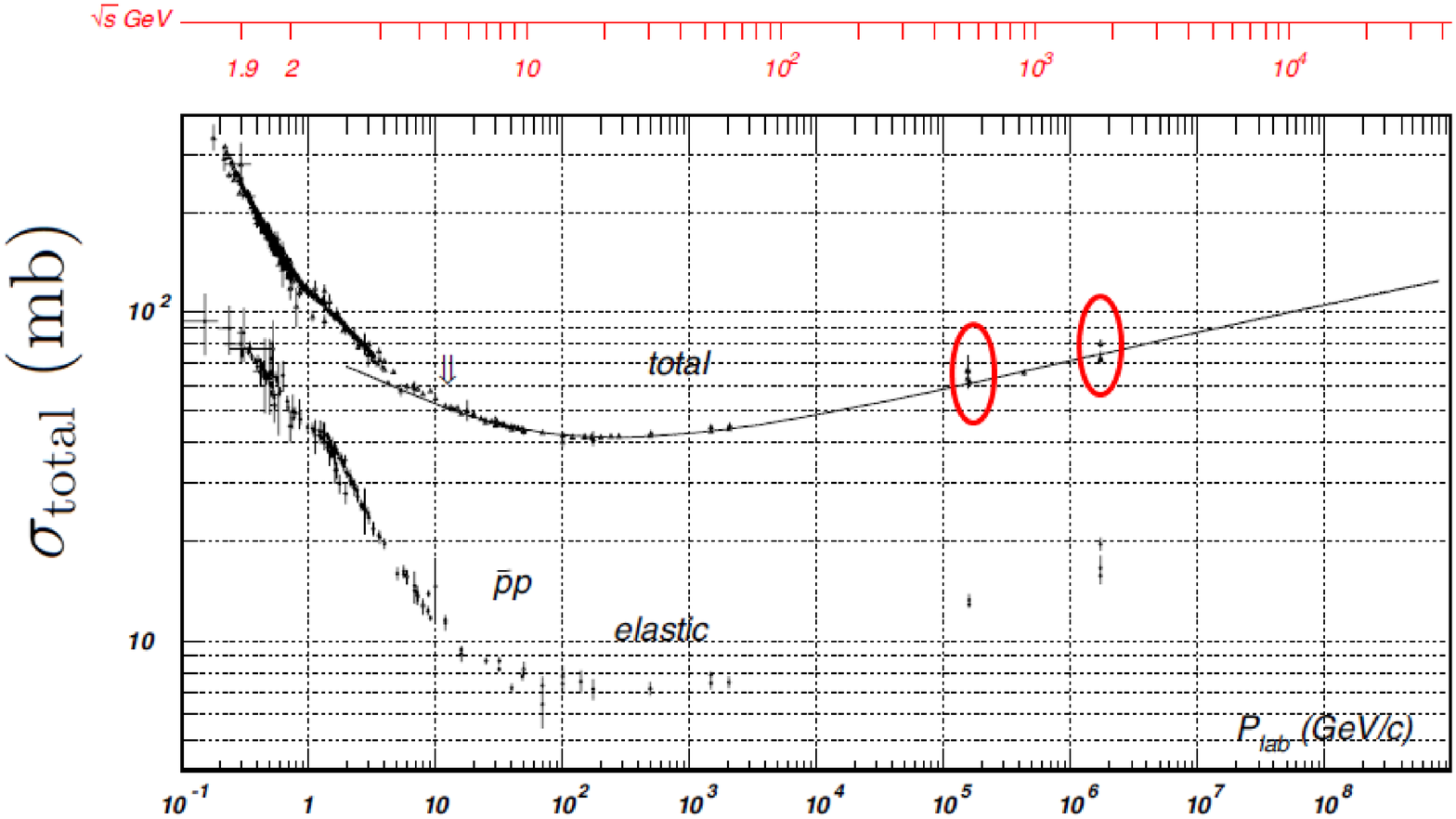} 
\includegraphics[width=8.0cm]{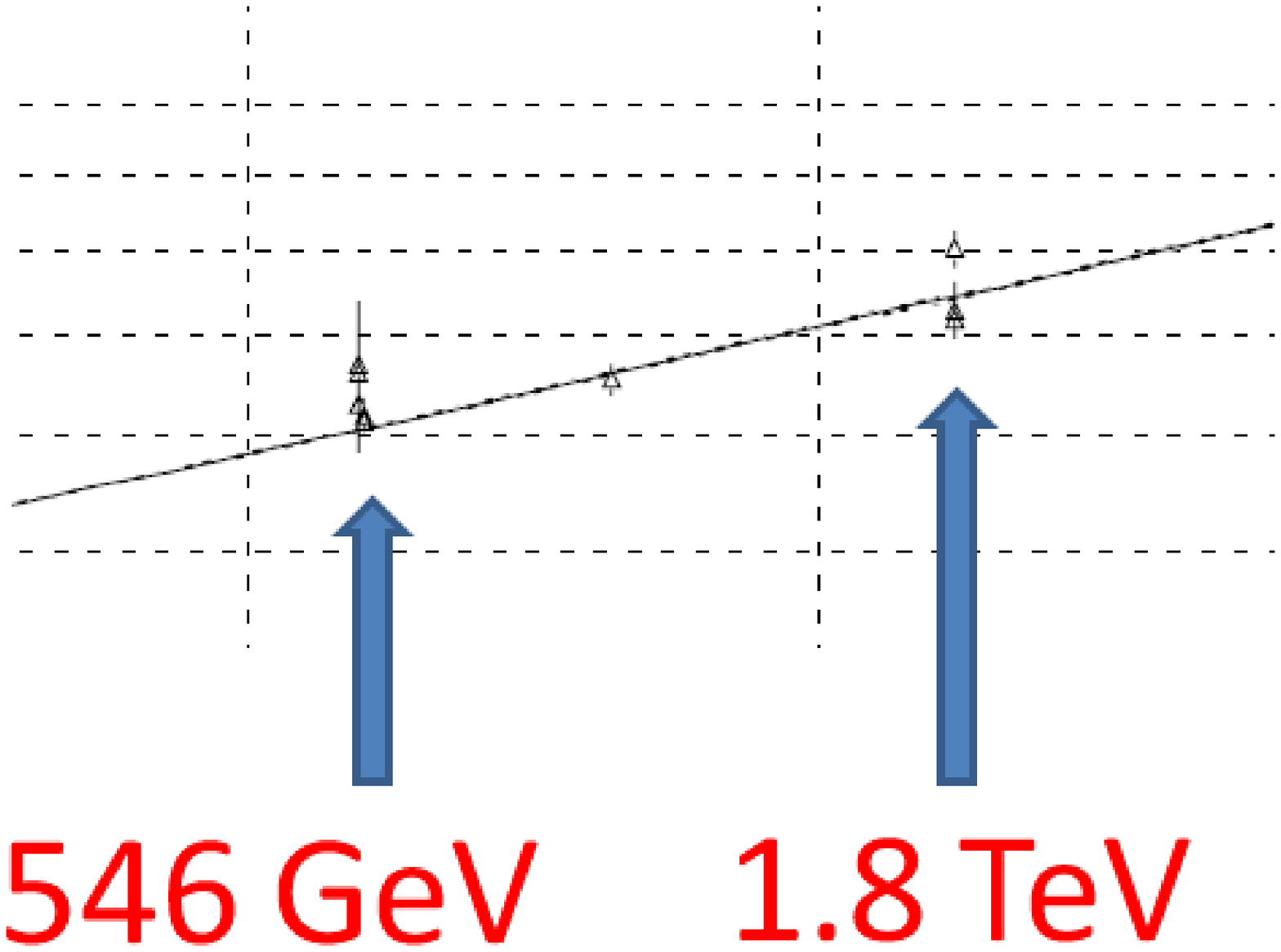} 
\caption{ \label{sigmapdg} Plots of total pp and $\rm p \bar p$ 
cross sections in the 
Review of Particle Properties of the Particle Data Group. }
\end{figure}

An important consequence of the precise description of $d\sigma/dt$ 
in terms of amplitudes valid for all $|t|$ is that we  
determine the total cross section and the slope parameters with 
precision  based on the whole set of data points in $t$.
In this way we can observe more clearly the origin of discrepancies 
 in reported values of $\sigma$ that occur,  
historically, at 541/546 GeV and at 1.8/1.96 TeV. This question is 
recalled in   Fig \ref{sigmapdg}, where parts of the plots in the 
Review of Particle Properties of the Particle Data Group \cite{PDG}
 are reproduced. 
We stress that in most cases  total cross sections  are not 
determined by direct measurements, but extracted from the measured 
elastic scattering distributions by some extrapolation procedures 
or model dependent calculations.
It must be remarked 
that the publications of the experimental groups have taken care to 
distinguish the experimental and calculated information in separate 
papers.

\begin{figure}[b]
\includegraphics[width=7.5cm]{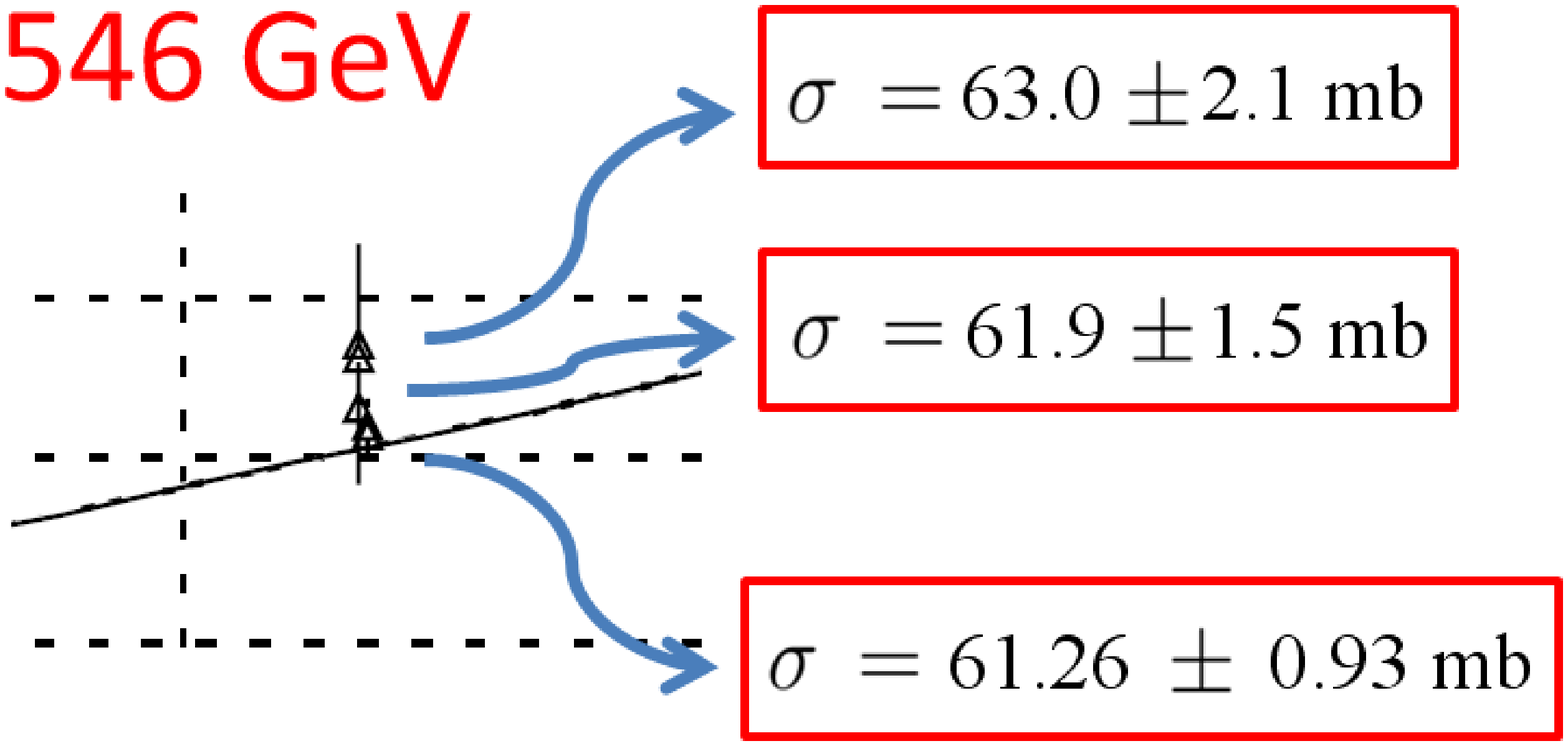}
\includegraphics[width=7.5cm]{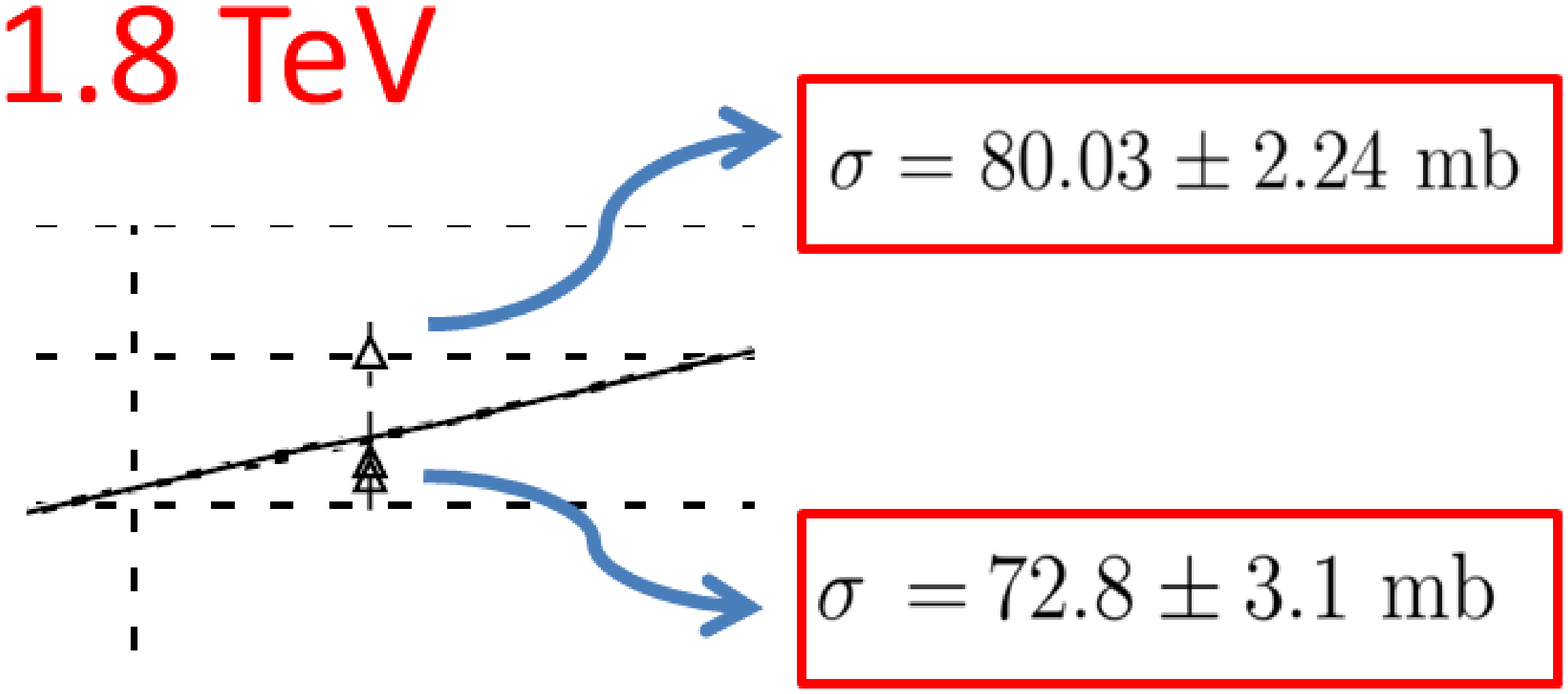} 
\caption{\label{1800GeV_figures} Variety of  values of total 
cross sections at 0.546 and 1.8 TeV  reported in literature \cite{PDG}.
We believe that the large error bars result from methods concentrating on 
forward scattering points, ignoring independence of slopes of amplitudes  
and using a single exponential form for $d\sigma/dt$. We stress that in 
general, total cross sections are not direct measurements, but rather 
are calculated values based on elastic scattering data. We thus 
remark  that a more complete treatment of all elastic scattering data 
is necessary, as done in our work.  } 
\end{figure}

The variety of reported values of cross sections at 0.546 and 1.8 TeV 
 are shown in Fig. \ref{1800GeV_figures}.

 \section{Analysis of data at high energies}

 \subsection{  $\sqrt{s}$ = 7 TeV } 

\begin{figure}[b]
\includegraphics[width=7.5cm]{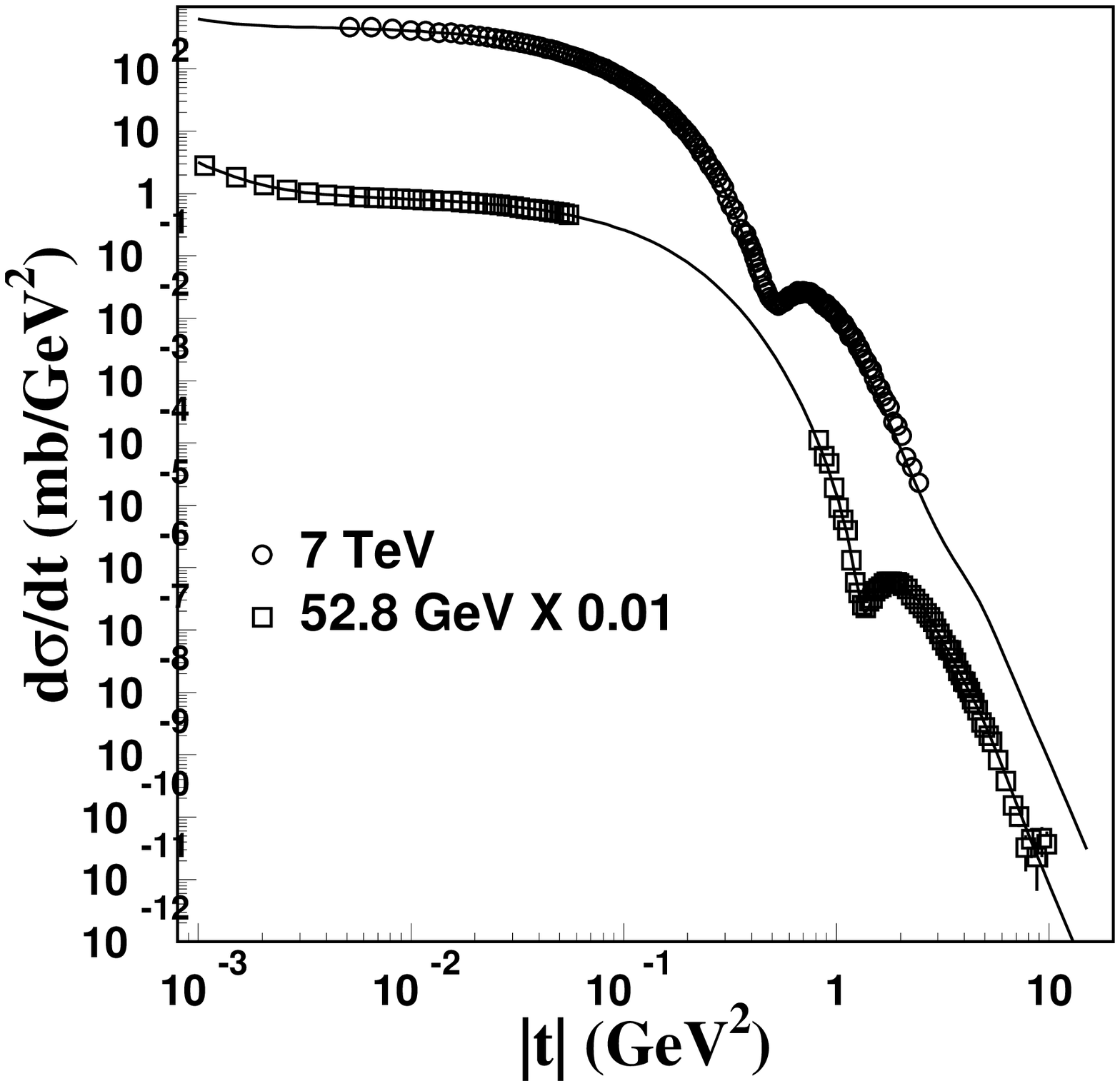} 
\includegraphics[width=7.5cm]{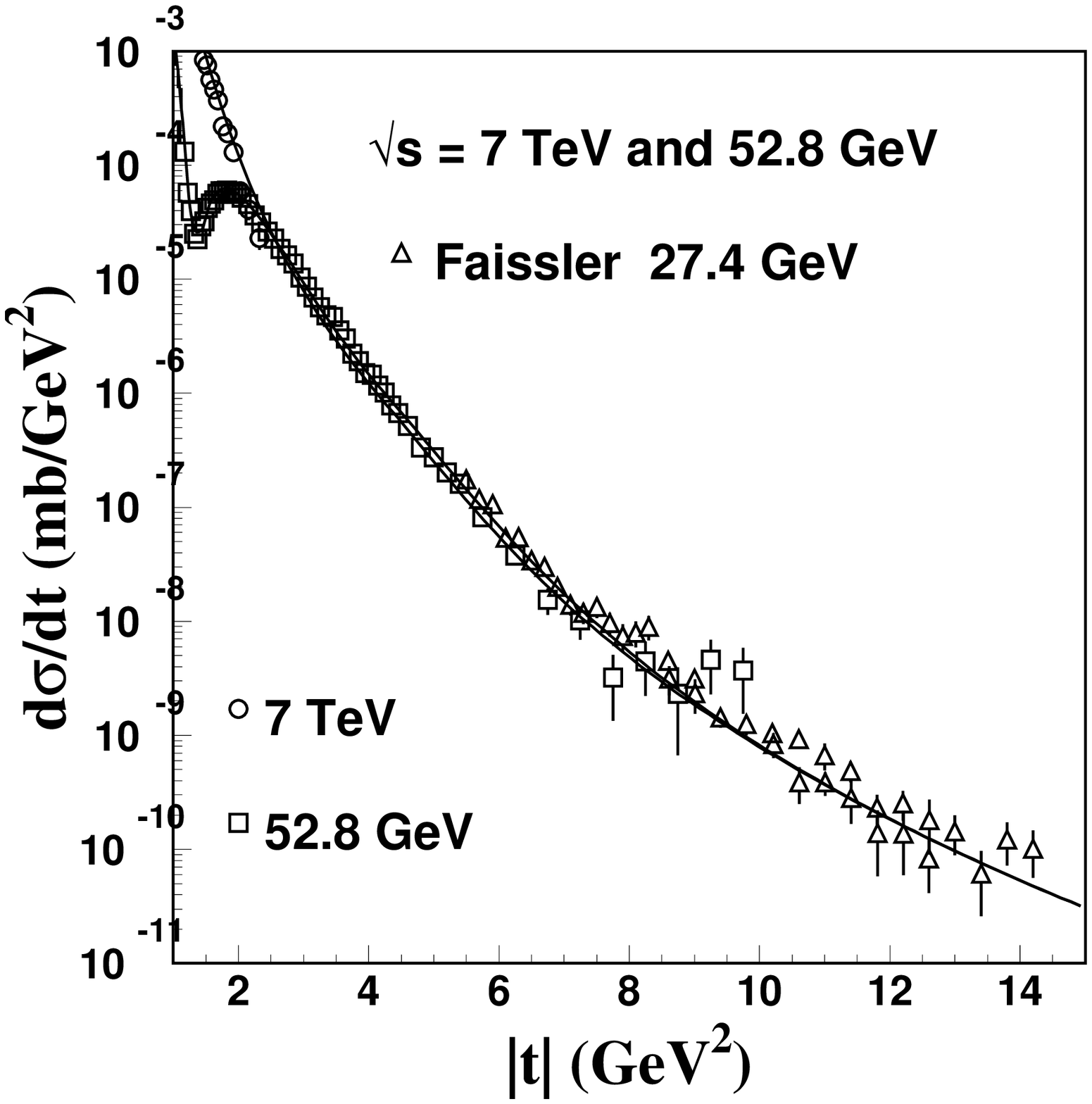} 
\caption{\label{7TeV_figures} Description of the data of $d\sigma/dt$ 
at 7 TeV and 52.8 GeV. The assumed connection with the universal tail at
large $|t|$ is pointed out in the second plot, where it looks like natural. }
\end{figure}

The fitting \cite{KEK_2013} of the 165 points of the TOTEM measurements 
at 7 TeV \cite{TOTEM}  is shown in Fig. \ref{7TeV_figures}. To show the 
regularity of the method, we plot together the results for  52.8 GeV. 
Parameters are given in the tables. 

The presence of a  perturbative tail for very large $|t|$ 
beyond  5 GeV$^2$ \cite{Faissler} is observed a 52.8 GeV and predicted 
for 7 TeV, as shown in  the second part of the same figure. 

\subsection{  $\sqrt{s}$ = 1.8/1.96 TeV }

\begin{figure}[b]
\includegraphics[width=7.5cm]{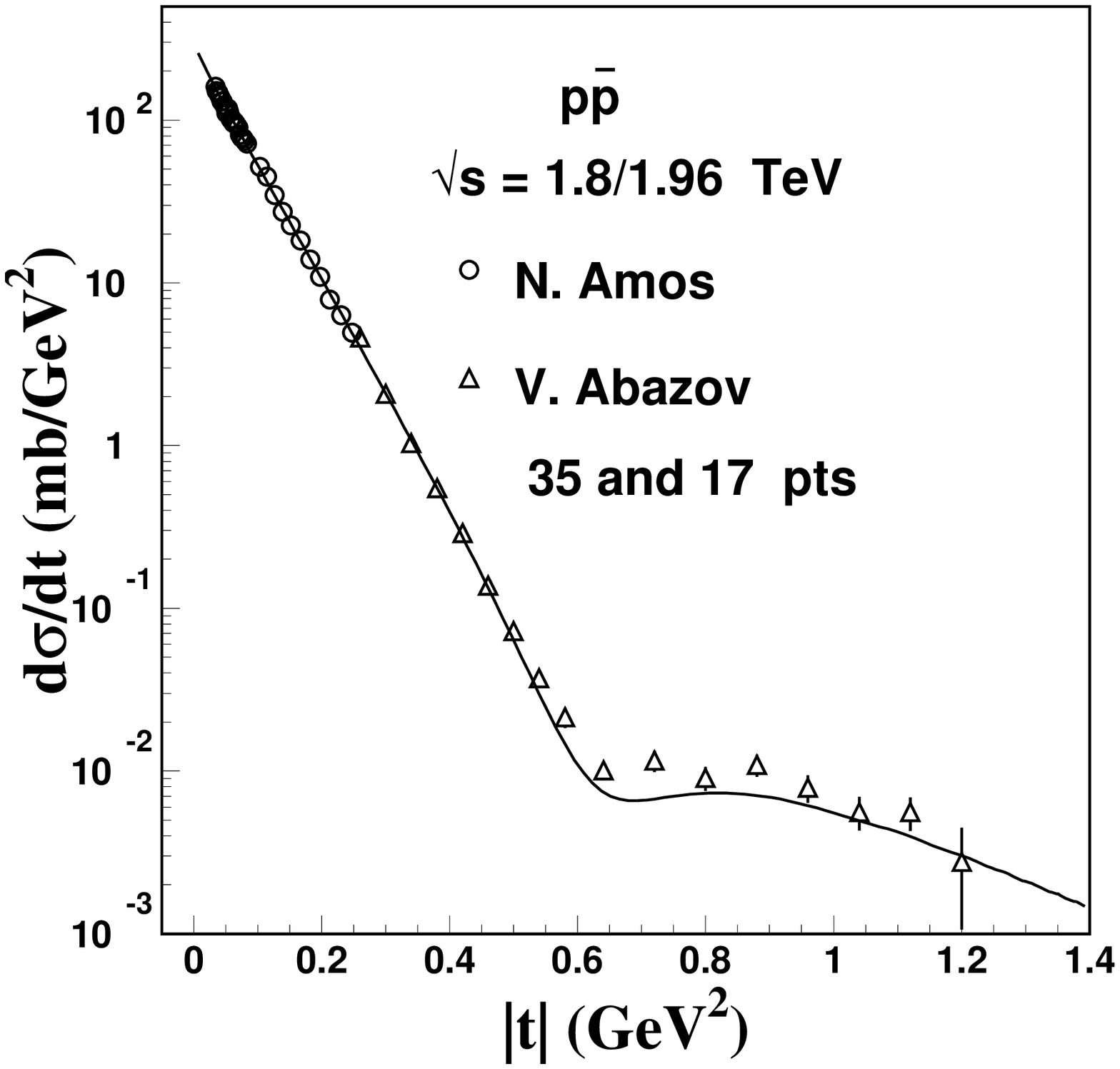} 
\includegraphics[width=7.5cm]{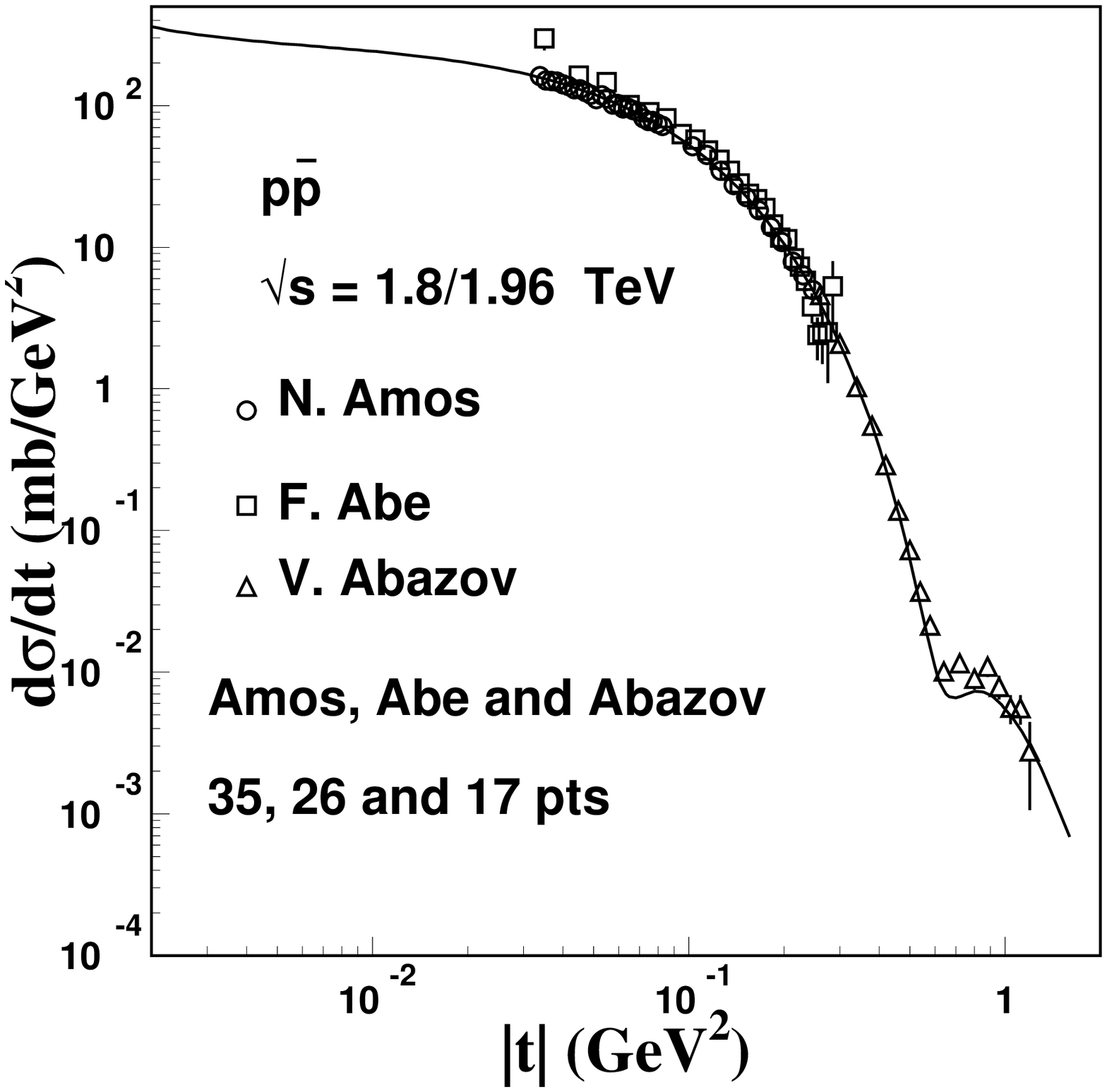} 
 \caption{ \label{cross_1800-figures} Data of the E-710 \cite{Amos},
CDF \cite{Abe} and D0 \cite{ppbar1960} experiments at 1.8/1.96  TeV. 
The line is a fit of a basis of 52 points formed with E-710 and D0 
data  \cite{KEK_2013b}. The parameters are given in the tables, 
and the amplitudes in $b$-space are plotted in Fig. \ref{b-figures}. } 
 \end{figure}

The  $d\sigma/dt$ data   of the E-710 \cite{Amos}, CDF \cite{Abe}  
and D0 \cite{ppbar1960} experiments  are shown in 
 Fig. \ref{cross_1800-figures}.
The line shown in the figure is a fitting  of a basis of 52 points
formed with the E-710 and D0 data.   Numerical values of the parameters 
of this line are given in the tables.

A complete analysis of all data at 1.8 and 1.96 TeV  
\cite{KEK_2013b} comparing  results obtained with different sets of 
data, taken from E-710,  CDF \cite{Abe} and  D0
experiments at Fermilab  suggests separate treatments in three 
different datasets, which reduces the discrepancies in values of 
the total cross  section, but still points to the need of  a judgment 
about the conditions of the two conflicting experiments made at 
Fermilab  in the first years of the 1990 decade. 

\subsection{ $\sqrt{s}$ = 0.546 TeV } 
  
\begin{figure}[ptb]
 \includegraphics[width=7.5cm]{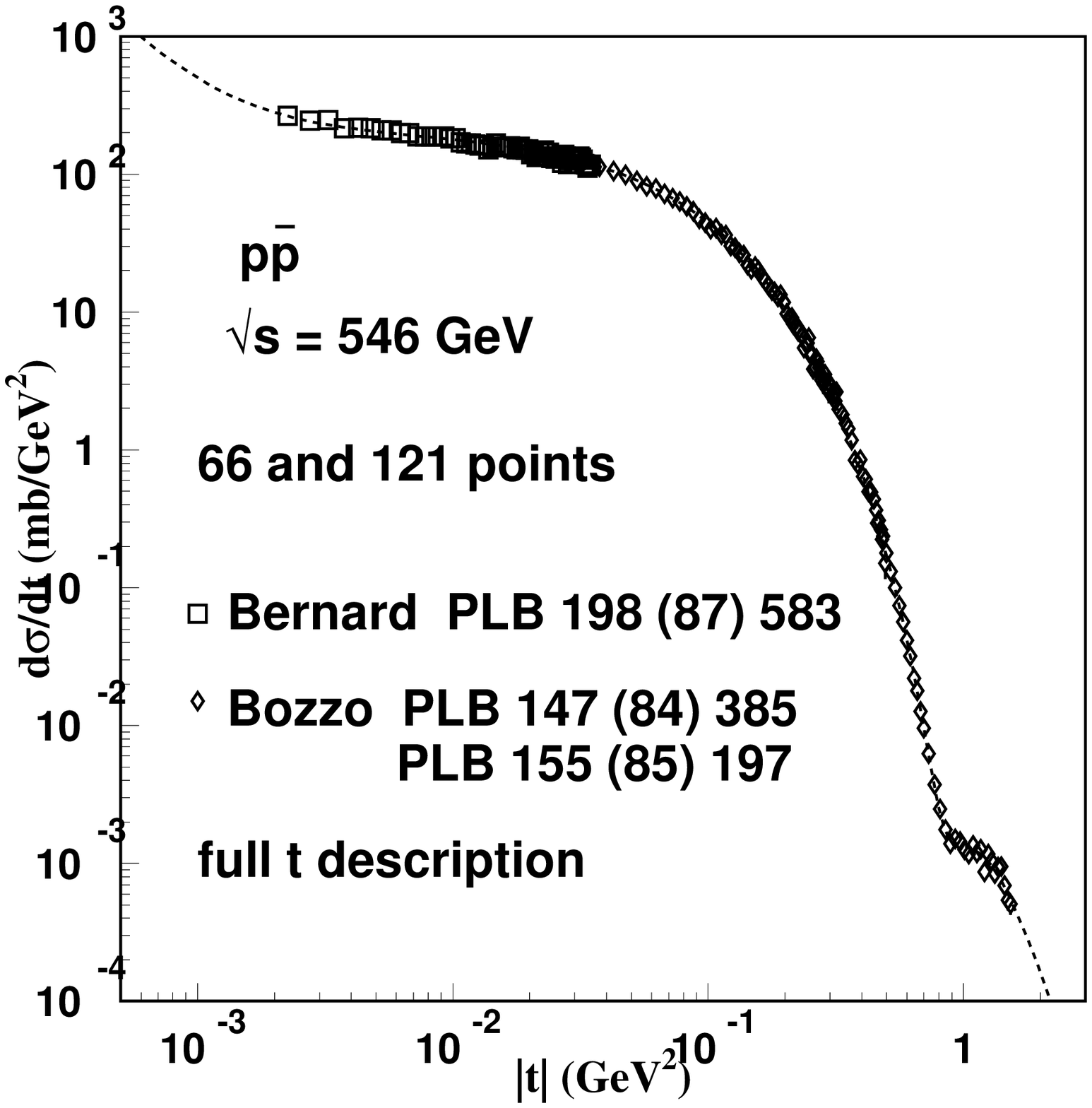}
\includegraphics[width=7.5cm]{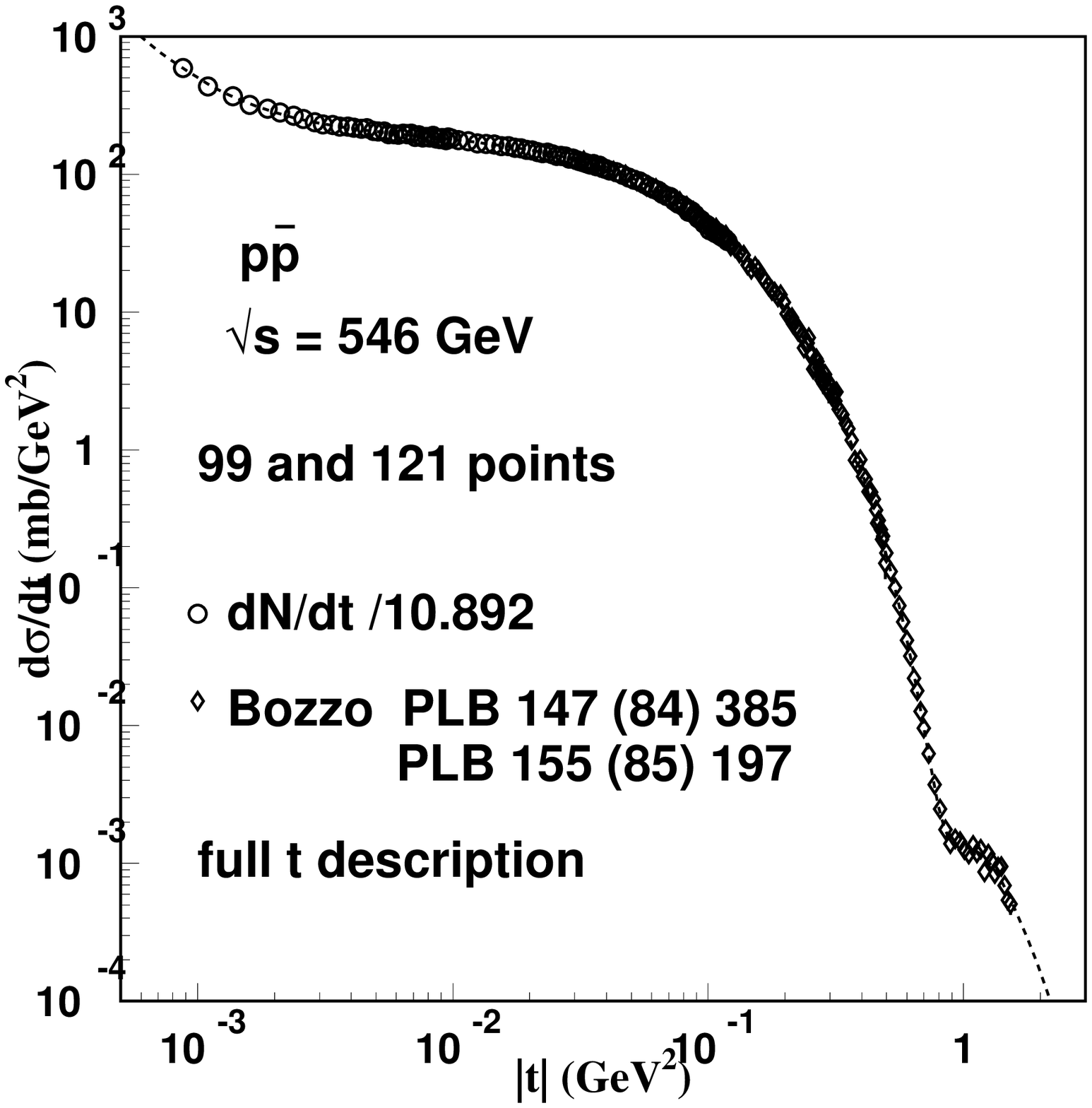} 
\caption{
\label{logplots} The low $|t| $ points of the Bernard et al. 
\cite{Bernard} and  event rate  \cite{Augier}  measurements,
put together with the points of Bozzo et al \cite{Bozzo1,Bozzo2,Bozzo3}
enrich the description of the differential cross section at 
546 GeV and allows a more precise determination of scattering 
parameters. The description of two different assemblages is consistent, 
with same parameter values, that are given in the tables.
The amplitudes in $b$-space   are plotted in Fig. \ref{b-figures}. 
}
\end{figure}

At 546 GeV the best quality data come from the UA4 experiment   
 \cite{Bozzo1,Bozzo2,Bozzo3}, with 121  points, which unfortunately do 
not reach the very low $|t|$ region. There are low $|t|$ data by 
Bernard et al. \cite{Bernard} and event rate points 
(not normalized)   by Augier et al. \cite{Augier}. We treat all these 
measurements in a unified way, with a description of high precision, 
shown in Fig. \ref{logplots}, where the log(t) plots are used to  enhance 
the forward region.  The numerical values are given in the tables.

 \begin{table*}[ptb]
\caption{ Free Independent Parameters  \vspace{0.1cm} } 
 \label{first_table} 
\tabcolsep=0.009cm
\begin{tabular}
[c]{|cc|cccc|cccc|c|}
%     \cline{3-10}
\hline
     &    &  \multicolumn{4}{c|}{imaginary amplitude } & \multicolumn{4}{c|}{real amplitude} &  \\
 \cline{3-10}
 $\sqrt{s}$   &  N   & $ \sigma  $ & $B_I$  &  $\alpha_I$ &  ~$\beta_{I} $ ~~ & $\rho$  &  $B_R$ &  $\lambda_{R} $ & 
      $\beta_R$  & $\langle\chi^{2}\rangle$\\
 TeV   & points & mb &     GeV$^{-2}$ & GeV$^{-2}$ & GeV$^{-2}$ &      &   GeV$^{-2}$   &   GeV$^{-2}$  & GeV$^{-2}$ & \\
\hline
~ 0.0307 ~ & ~  201 ~ & ~ 40.38 ~ & ~ 12.47 ~ & ~ 6.0180 ~ & ~ 2.3437 ~ & ~ 0.036 ~ & ~ 25.0 ~ & ~ 0.5226 ~ & 
   ~ 0.8 ~  & ~ 2.430 ~   \\ 
~ 0.0446 ~ & ~  230 ~ & ~ 41.84 ~ & ~ 13.10 ~ & ~ 6.0953 ~ & ~ 2.4321 ~ & ~ 0.0545 ~ & ~ 18.0 ~ & ~ 0.7823 ~ & 
   ~ 0.96 ~  & ~ 1.940 ~   \\ 
0.0528  & ~  97 ~ & ~ 42.49  ~ & ~ 13.04 ~ & ~ 5.9561 ~ & ~ 2.3477 ~ & ~ 0.078 ~ & ~ 19.07 ~ & ~ 1.1307 ~ & 
   ~ 1.1436  ~  & ~ 0.8328 ~   \\ 
0.0625  & ~  138 ~ & ~ 43.25 ~ & ~ 13.20 ~ & ~ 5.9202 ~ & ~ 2.3336 ~ & ~ 0.086 ~ & ~ 19.0 ~ & ~ 1.2480 ~ & 
   ~ 1.3282 ~  & ~ 2.284 ~   \\ 
0.546   & ~  180 ~ & ~ 62.00 ~ & ~ 15.37 ~ & ~ 10.363  ~ & ~ 3.4336  ~ & ~ 0.1350  ~ & ~ 22.00 ~ & ~ 2.8849 ~ & 
   ~ 1.00  ~  & ~ 1.0807 ~   \\ 
0.9  & ~  ~ & ~ 65.30 ~ & ~ 16.0 ~ & ~10.80 ~ & ~ 3.6 ~ & ~ 0.138 ~ & ~ 23.0 ~ & ~ 3.13 ~ & 
   ~ 1.05 ~  & ~   ~   \\ 
1.8   & ~ 52  ~ & ~ 72.76 ~ & ~ 16.76 ~ & ~ 11.62 ~ & ~ 3.7633 ~ & ~ 0.1415  ~ & ~ 24.00 ~ & ~ 3.5400 ~ & 
   ~ 1.15 ~  & ~ 1.3412 ~   \\ 
7  & ~ 165  ~ & ~ 98.65 ~ & ~ 19.77 ~ & ~ 13.730 ~ & ~ 4.0826  ~ & ~ 0.141 ~ & ~ 30.20 ~ & ~ 4.7525 ~ & 
   ~ 1.4851  ~  & ~ 0.3105  ~   \\ 
14  & ~   ~ & ~ 108~ & ~ 21.0 ~ & ~ 14.4 ~ & ~ 4.3 ~ & ~ 0.141 ~ & ~ 32.0 ~ & ~ 5.17 ~ & 
   ~ 1.50 ~  & ~   ~   \\ 
57  & ~   ~ & ~ 131 ~ & ~ 22.0 ~ & ~ 16.3 ~ & ~ 4.7 ~ & ~ 0.141 ~ & ~ 35.0 ~ & ~6.3 ~ & 
   ~ 1.6 ~  & ~  ~   \\ 
\hline 
\end{tabular}
\end{table*}
%$\langle\chi^2\rangle$
\begin{table*}[ptb]
\caption{ Derived properties of amplitudes and cross sections \vspace{0.1cm} }  
\label{second_table}
%\begin{center}
%\tabcolsep=0.01cm
\begin{tabular}
[c]{|ccccccccccccc|}\hline
$\sqrt{s}$ &  $\mathrm{Z_{I}} $ ~ & $\mathrm{Z_{R}(1)} $~ &  $\mathrm{Z_{R}(2) }$~ &  $|t|_{\mathrm{dip}}$ &
            $|t|_{\mathrm{bump}}$ & ratio  & $T_I(b=0)$ & $T_R(b=0) $ & $\sigma$(el,Real)& $\sigma$(el)& $\sigma$(inel)& 
                       $\sigma$(el)/$\sigma$\\
  TeV & GeV$^{2}$ & GeV$^{2} $ & GeV$^{2}$ & GeV$^{2}$ & GeV$^{2}$ &   & GeV$^{2}$ & GeV$^{2}$ &   mb  & mb & mb &  \\
 \hline
0.0307 &1.4016 & 0.1828 & 1.6079 & 1.4039 & 1.8794 & 79.95 &1.2839 & 0.0028 & 0.004  &  7.070 & 33.31 & 0.1751  \\
\hline
0.0447 &1.3364 & 0.2874& 1.6607&1.3733 &1.8107 & 3.710 & 1.2531 & 0.0232& 0.014 & 7.183& 34.66 & 0.1717  \\
\hline
 0.0528 & 1.3083 & 0.2710 & 1.6157 & 1.3560 & 1.7947  & 3.281 & 1.2685 & 0.0310 & 0.028  & 7.431 & 35.06 & 0.1749 \\
\hline
 0.0625 &  1.2794 & 0.2797 & 1.6201 & 1.3365& 1.7562 & 2.606 & 1.2685 & 0.0380 & 0.036 &7.593 & 35.66 & 0.1756 \\
\hline
 0.546 & 0.8282 & 0.2282 & 1.2551 & 1.0005 & 1.0005 & 1 & 1.5091 & 0.0750 & 0.156 & 13.402 & 48.60 & 0.2162 \\
\hline
 0.9 & 0.7425& 0.2170  & 1.1970     & 0.8897  & 0.8897 & 1 & 1.5000 & 0.0768 & 0.175 & 14.210 & 51.09 & 0.2176 \\
\hline
 1.8 & 0.6336 & 0.2080 & 1.1337 & 0.6946 & 0.8133 & 1.079 & 1.5439 &  0.0840  & 0.216 & 16.694 & 56.07 & 0.2294 \\
\hline
 7  & 0.4671 & 0.1641 & 0.8235 & 0.4847 & 0.6488  & 1.838 & 1.6815 & 0.0980 & 0.311  &25.542& 35.06 & 0.2589 \\
\hline
 14 & 0.4251 & 0.1568 & 0.7097 & 0.4371 & 0.5999  & 2.395 & 1.7093 & 0.1172 & 0.350 & 28.70 &  79.30 & 0.2657 \\
\hline
57  & 0.3285 & 0.1411 & 0.6608 & 0.3314 & 0.5019 & 6.172 & 1.7340& 0.1242 & 0.470 & 39.53 & 91.47 & 0.3017 \\
\hline
\end{tabular}
%\end{center}
\end{table*}
\begin{figure*}[b]
\includegraphics[height=5.8cm]{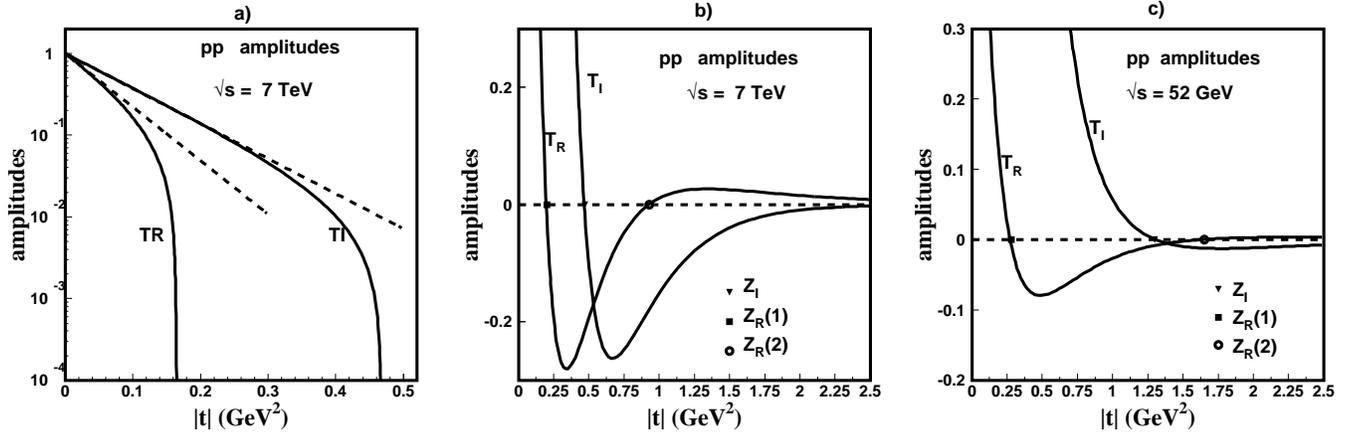} 
\caption{\label{amplitudes-figures} (a) Forward scattering
amplitudes $T_{R}$ and $T_{I}$ at $\sqrt{s} = 7 $ TeV in log scale, normalized
to one at $|t|=0$, showing their slopes, $B_{I}=19.77 $ and $B_{R}=30.2
\nobreak\,\mbox{GeV}^{-2}$, and their curvatures, and indicating positions of
the first zeros $Z_{R}(1)$ and $Z_{I}$; 
(b) Large-t dependence of the real and
imaginary scattering amplitudes showing the complete set of zeros; 
(c) t dependence of the real and imaginary scattering amplitudes at
$\sqrt{s}= 52.8$ GeV. Comparing with the figure for 7 TeV, we observe that 
all zeros  move  towards smaller  $|t|$ values as the energy increases. } 
\end{figure*}

\begin{figure*}[b]
\includegraphics[width=7.5cm]{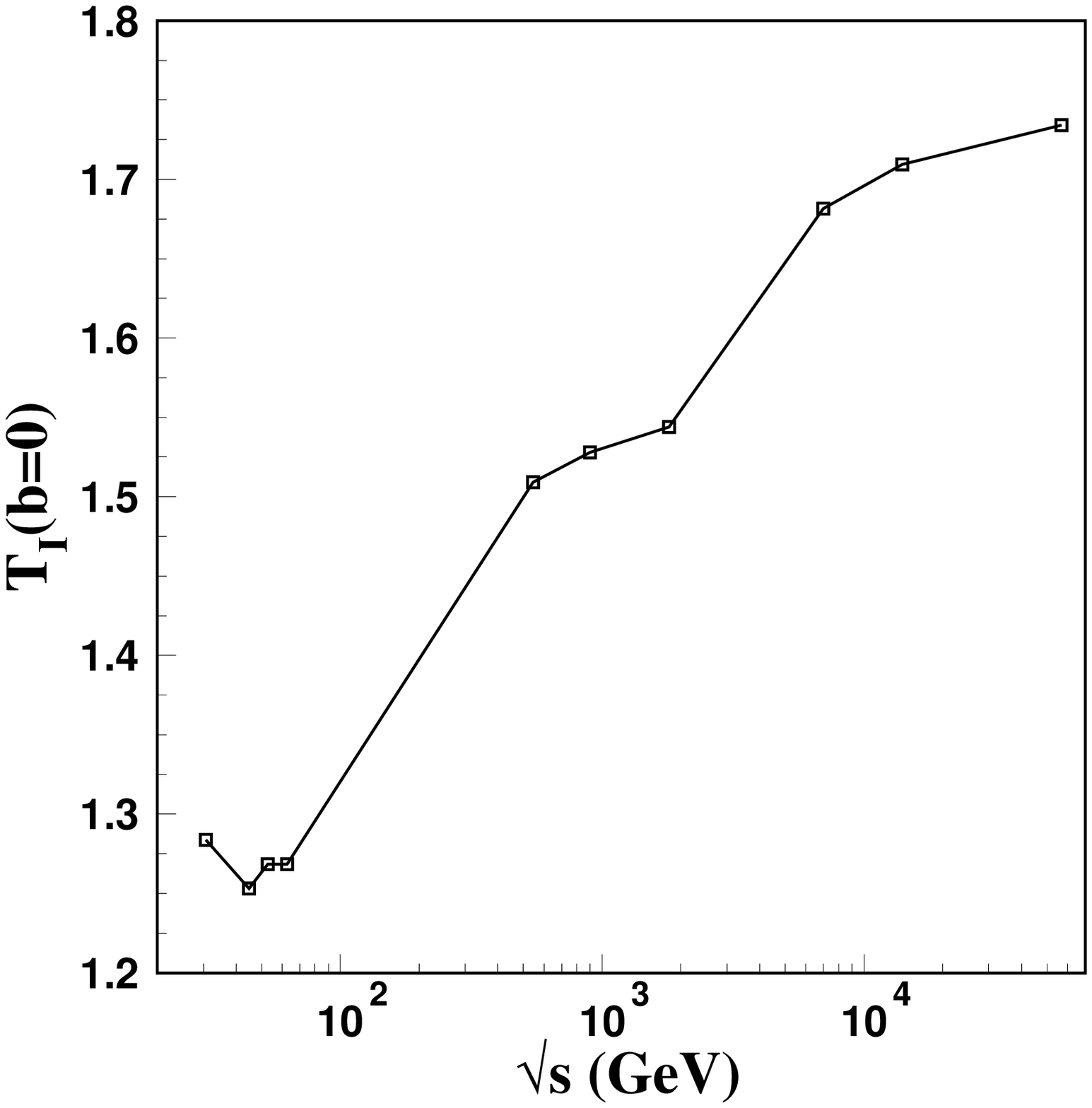} 
\includegraphics[width=7.5cm]{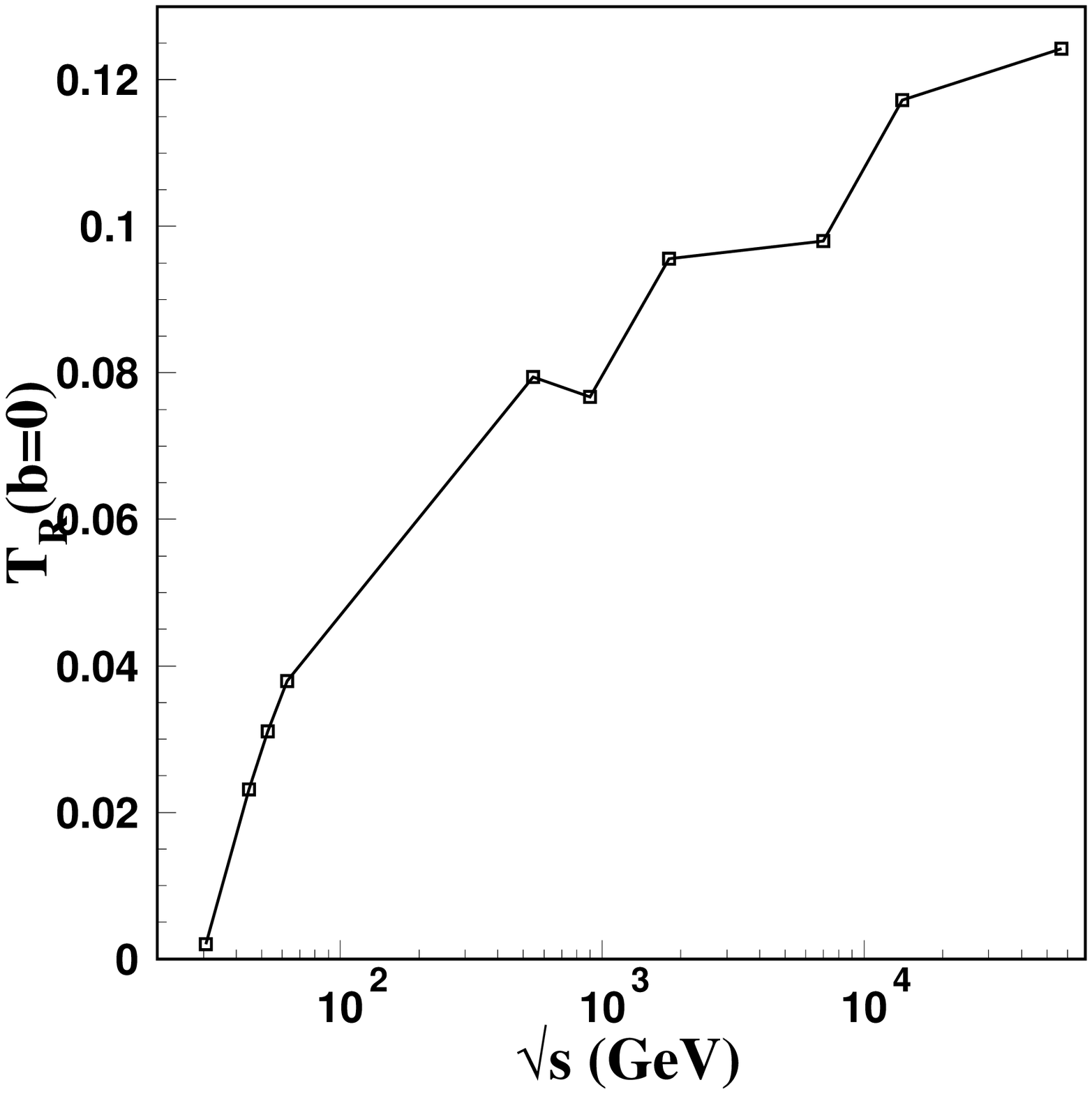} 
\caption{ \label{b-equal-zero} Values of the imaginary and real profile functions 
at $b=0$.  }
\end{figure*}

\begin{figure*}[b]
\includegraphics[width=7.5cm]{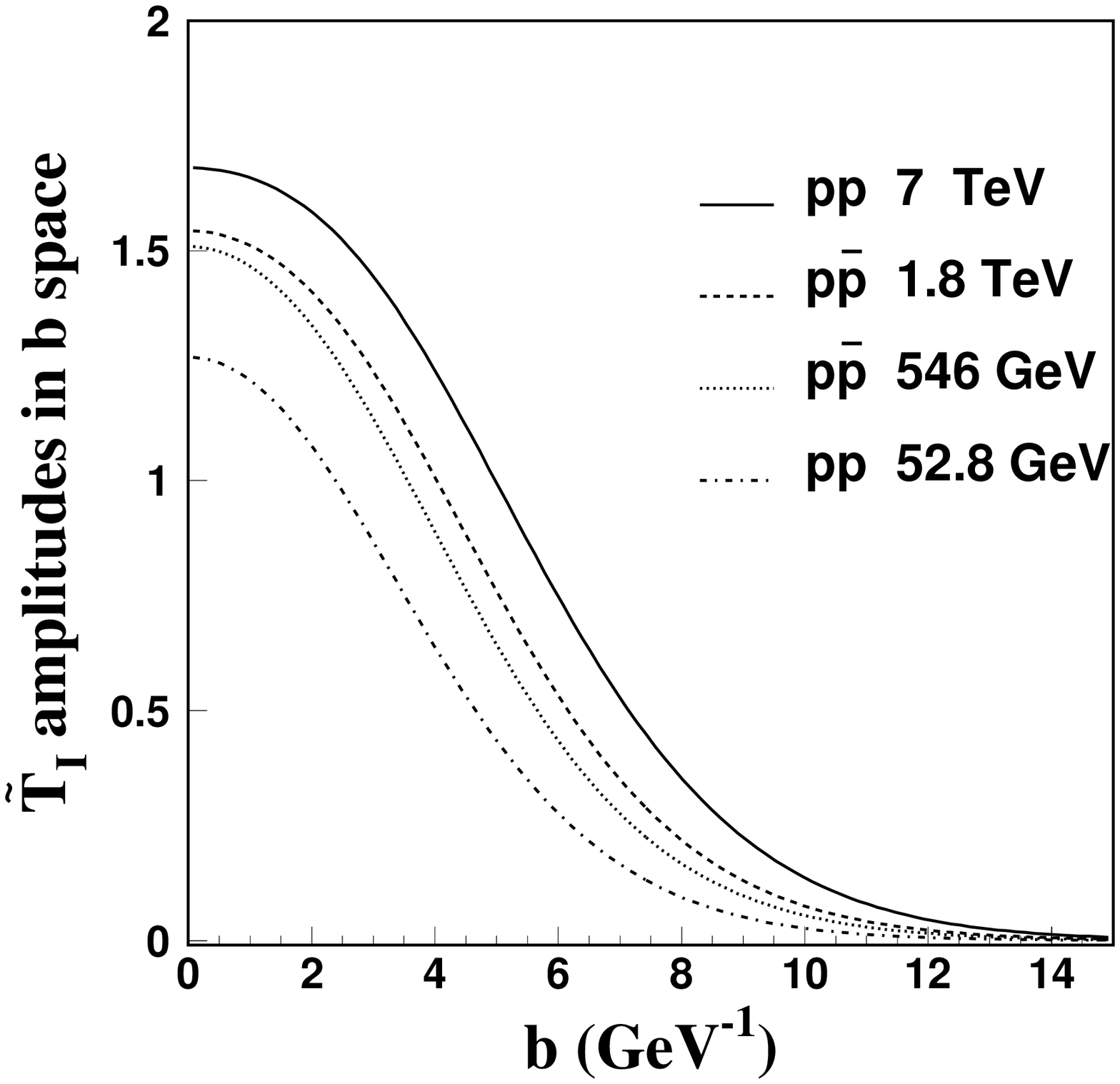} 
\includegraphics[width=7.5cm]{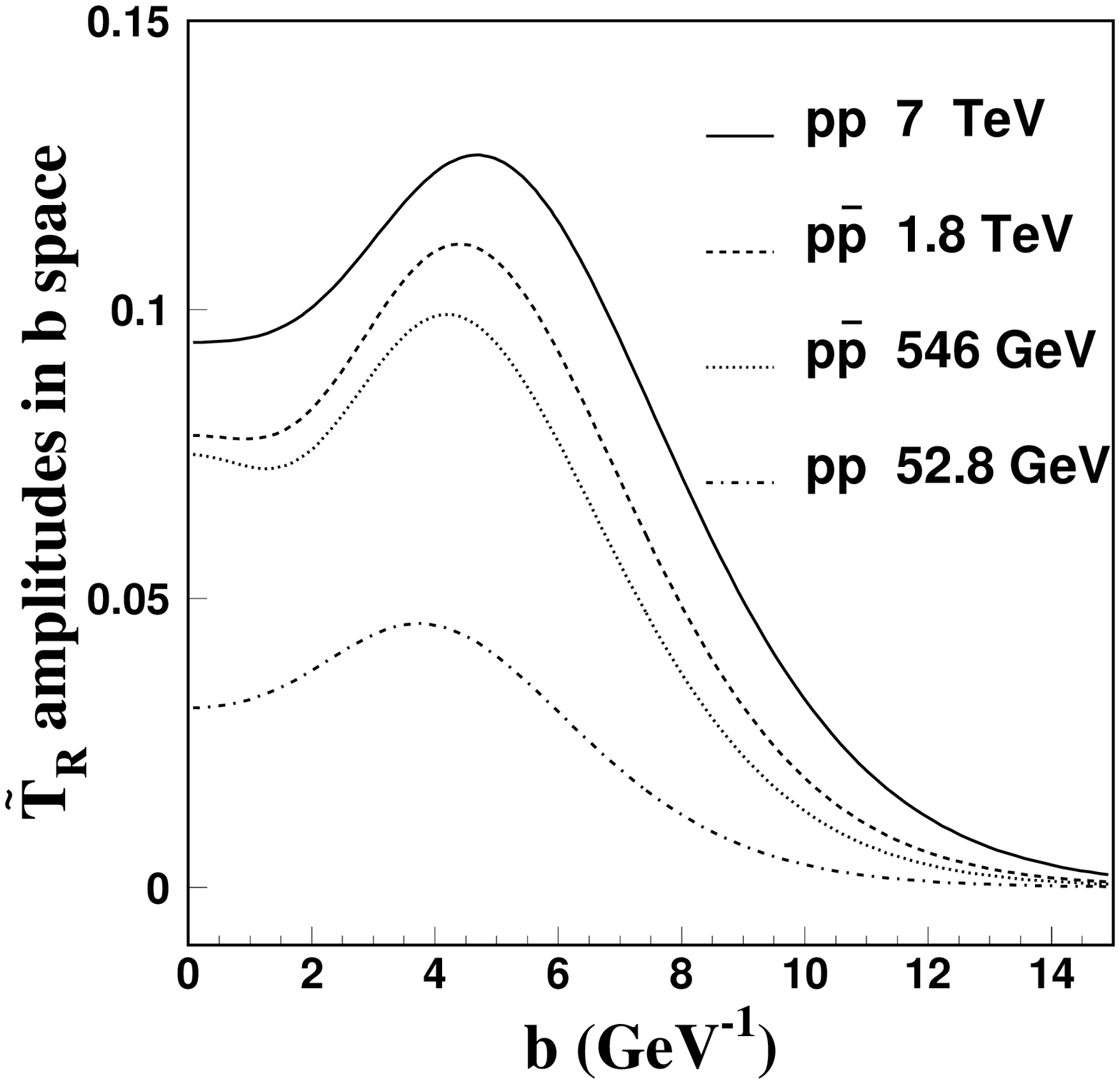} 
\caption{ \label{b-figures} Imaginary and real amplitudes in impact 
parameter space, for several energies. The regularity of the 
evolution of the functions as the energy varies is impressive.}
\end{figure*}

\begin{figure*}[b]
\includegraphics[width=7.5cm]{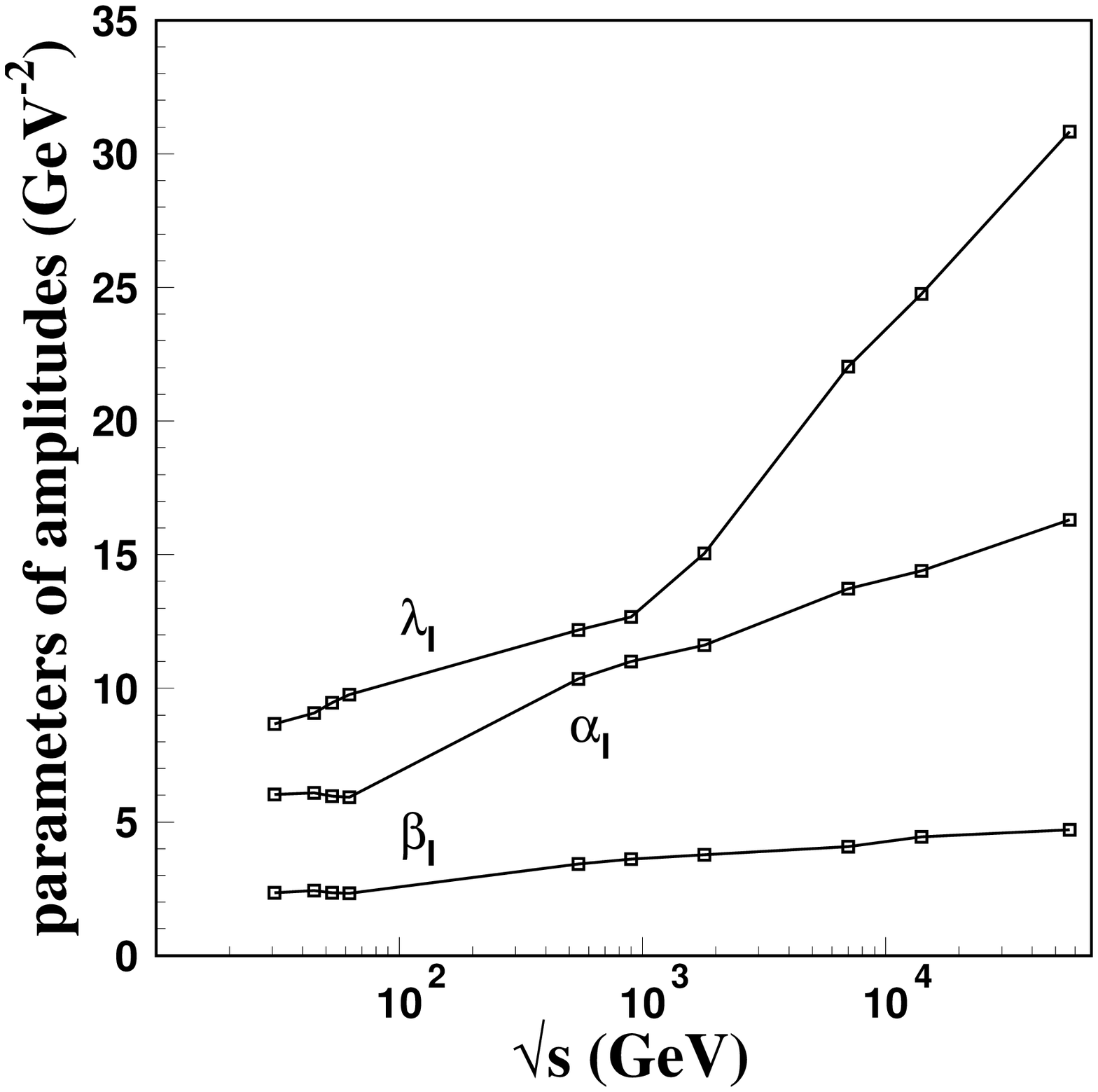} 
\includegraphics[width=7.5cm]{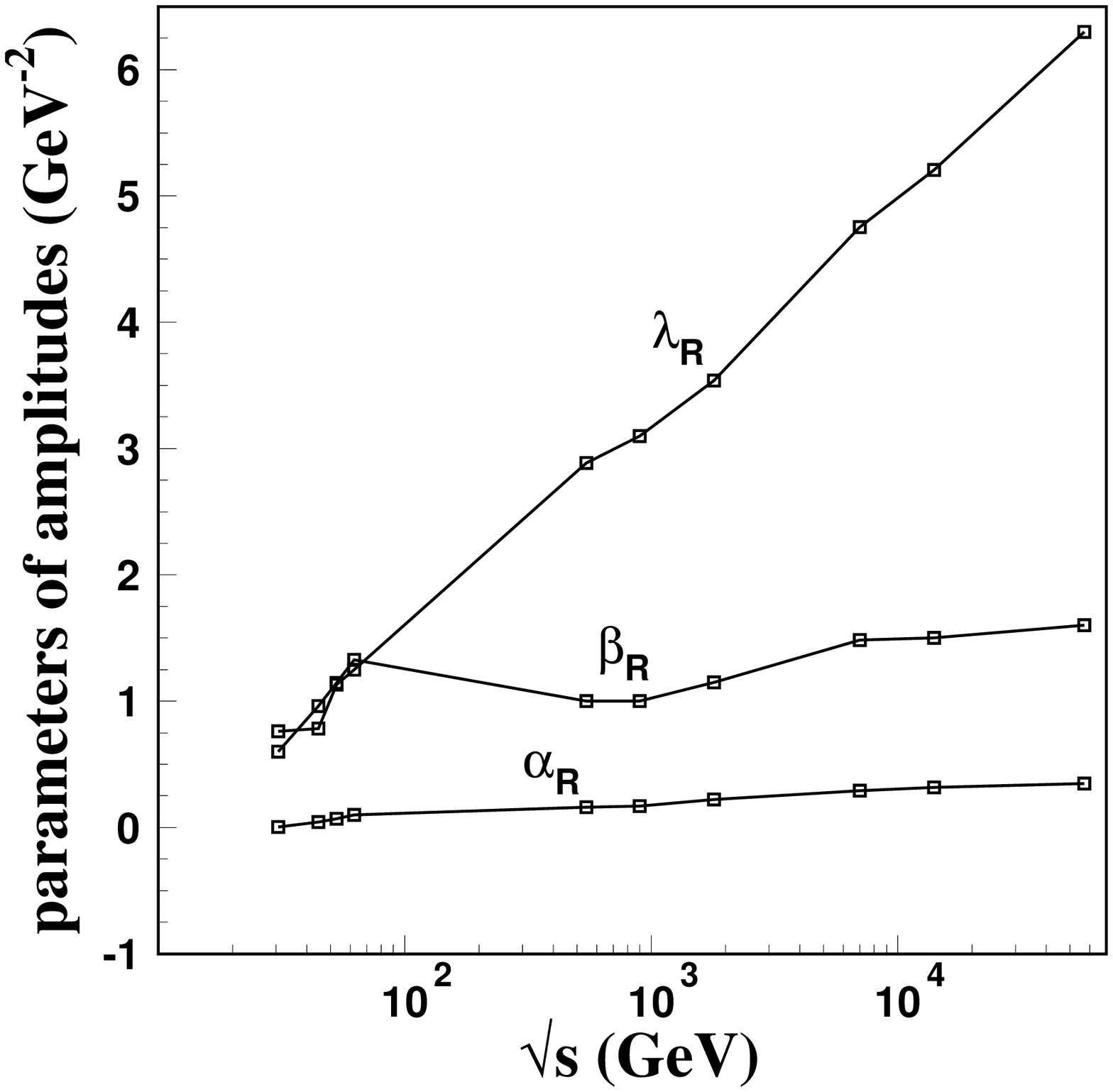} 
\caption{\label{parameters} Parameters of the amplitudes at the 
energies of the data. In each case, the fourth free parameter is 
the corresponding slope, $B_I$ or $B_R$, shown in the next figure.
 The sum $\alpha_I+\lambda_I$ gives the total cross section in 
$\GeV^{-2}$, 
and the sum $\alpha_R+\lambda_R$ gives the real amplitude at 
$|t|=0$, equal to $\rho \times \sigma$.  }  
 \end{figure*}

\begin{figure*}[b]
\includegraphics[width=7.0cm]{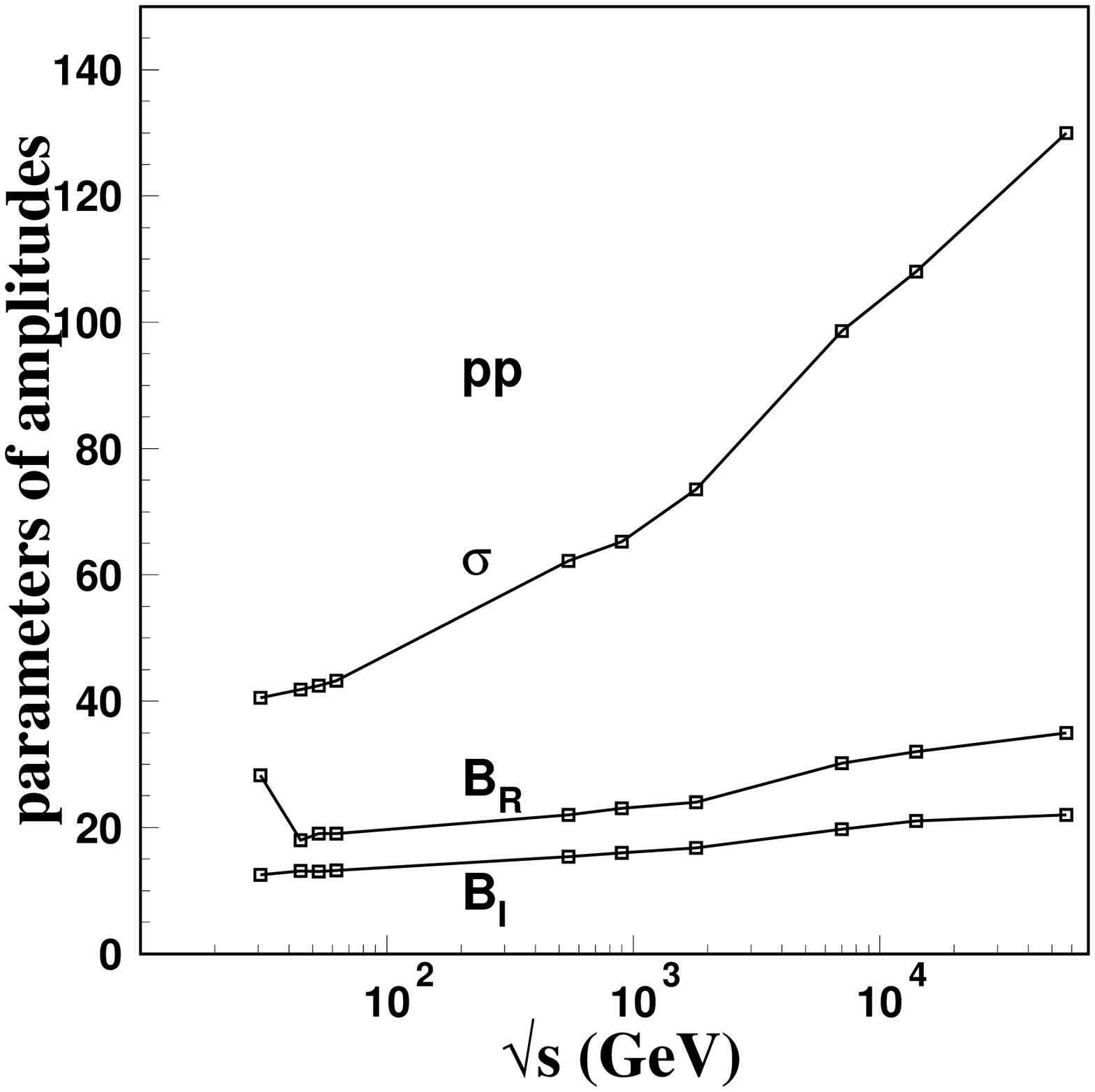} 
\includegraphics[width=7.0cm]{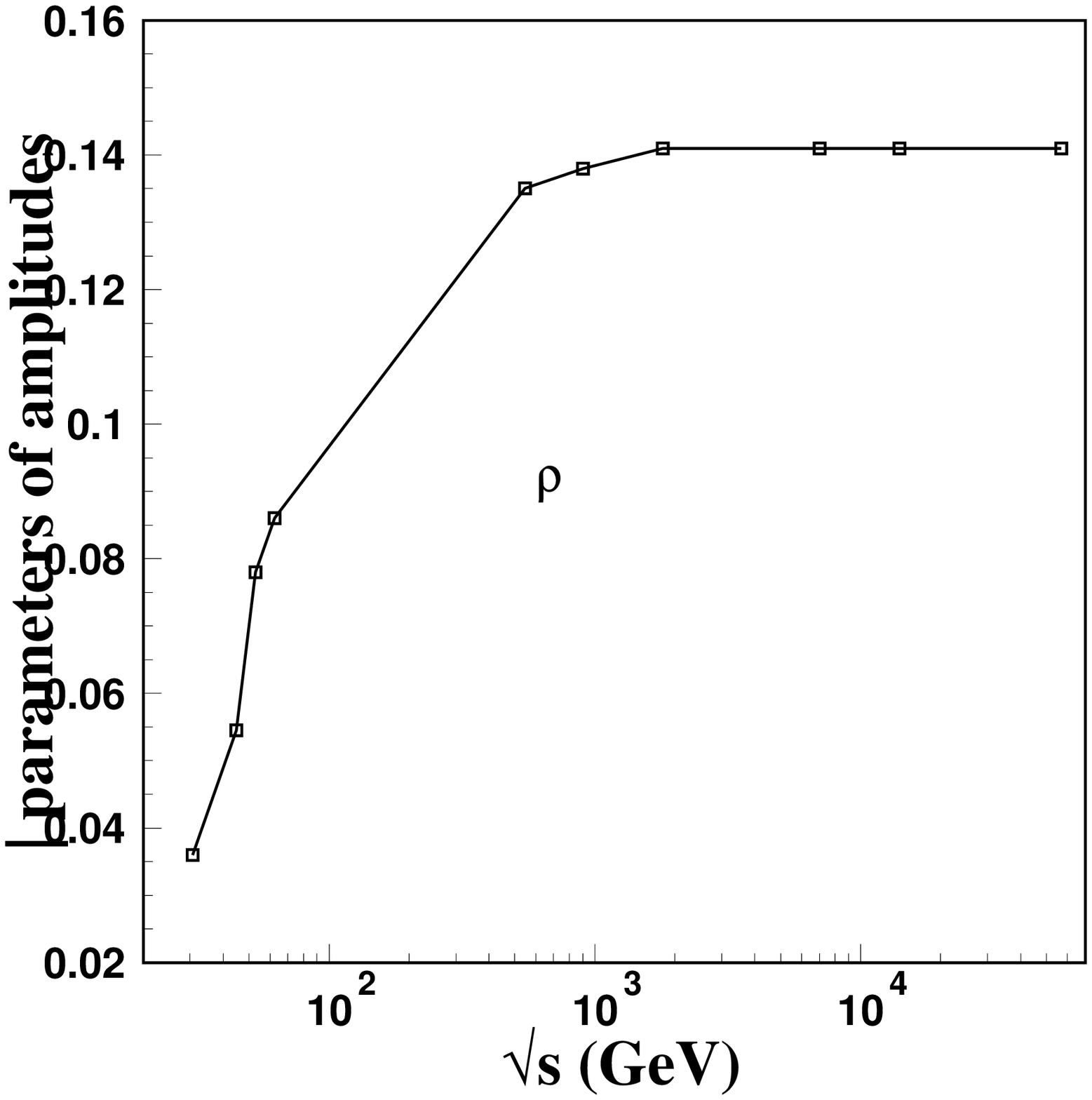} 
\caption{\label{forward-figures} Total cross-section and 
slopes characteristic of forward scattering; the $\rho$ parameter.}  
 \end{figure*}

\begin{figure}[b]
\includegraphics[width=7.0cm]{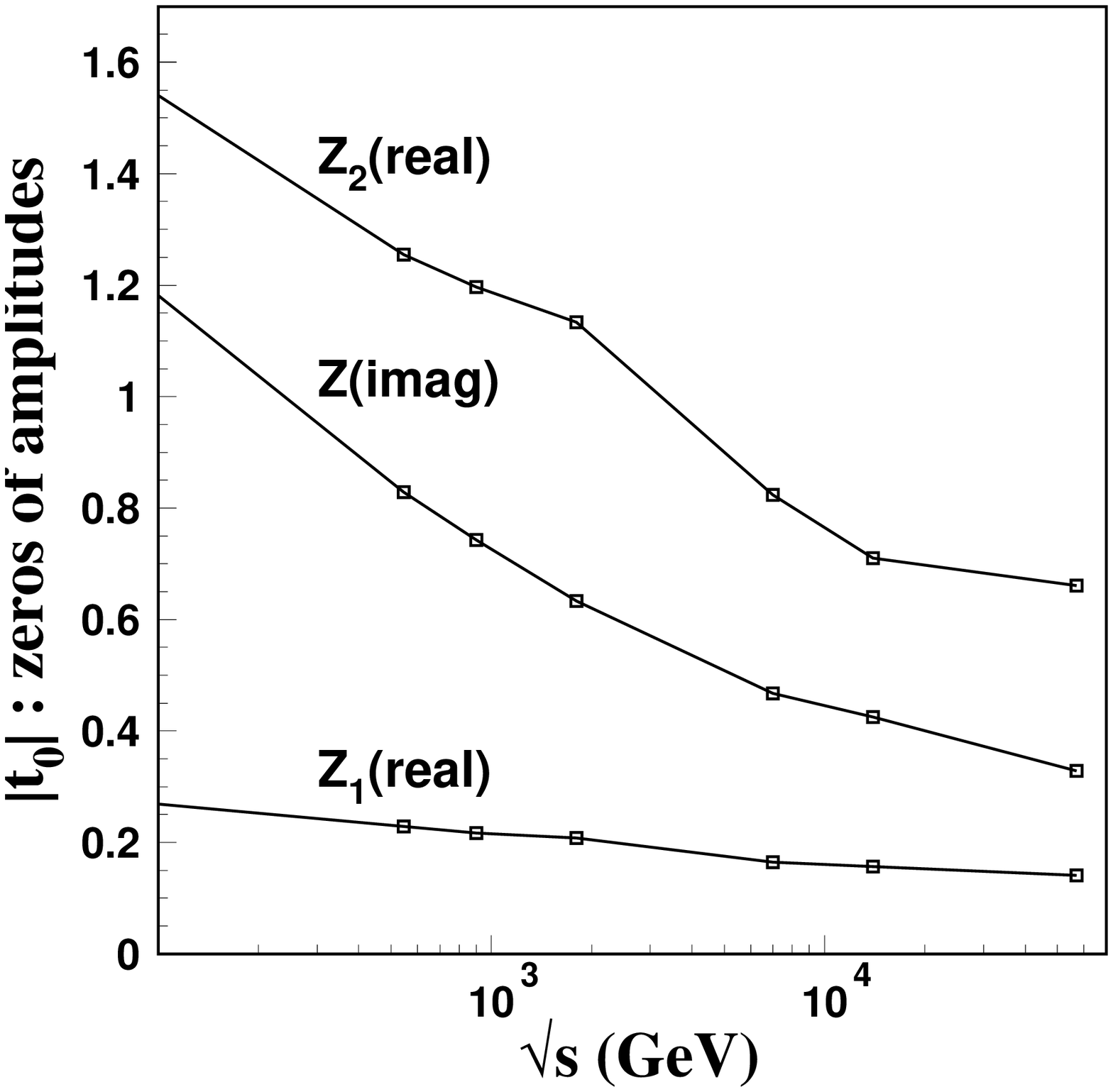} 
\caption{\label{forward-figures} Positions of 
zeros of the amplitudes.}  
 \end{figure}

\section{Amplitudes and Profile Functions }

The forms of the amplitudes are exemplified 
for the 7 TeV case \cite{KEK_2013} in Fig.\ref{amplitudes-figures}. 

We call attention for the important difference in slopes
of $T_I$ and $T_R$ and to the features that are general in our solution 
for the disentanglement: one zero for the imaginary part, and two zeros 
for the real part in pp scattering. In the  $\rm p \bar p$ case there may 
be a third real zero due to the negative contribution of the 
perturbative tail.  

It is important to observe that for large $|t|$ the real and imaginary 
amplitudes (nuclear part) decrease, with positive and negative signs 
respectively.

See also Fig. \ref{b-equal-zero}

The amplitudes in b-space for several energies are shown if 
Fig. \ref{b-figures}.
 Both $\tilde T_I(s,b)$ and $\tilde T_R(s,b)$ are always positive and fall to zero 
at large $b$ more slowly than Gaussians. The magnitudes of $\tilde T_R$
are always smaller than those of $\tilde T_I$ , differently from what happens 
in $t$ space, where there  are zeros and variations in magnitudes.

The integrated elastic cross section from the real part is much 
smaller than the corresponding value for the imaginary part, but
the its presence is crucial for the design of the $d\sigma/dt$ 
form at intermediate and large $t$ values.

\section{ Tables and Energy dependence }

The very regular behavior observed in the plots of the amplitudes
in Fig. \ref{b-figures}
is seen also in their characteristic parameters. This is shown 
in Figs. \ref{parameters} and \ref{forward-figures}.

The quantities characteristic of the forward range (total cross-section,
$\rho$ parameter and slopes) for the energies of the data, are shown
in Fig. \ref{forward-figures}. The extrapolations to the very high 
energy of the Auger cosmic rays experiment are interesting, 
giving total cross section value compatible with the 
evaluation by the experimental group \cite{Auger}. 

In our scheme, at all energies the imaginary part has only one zero. The 
second zero  of the real amplitude has important role in  the behavior 
of $d\sigma/dt$ at intermediate and large $|t|$ values.

\section{ Summary and Perspectives }

    In this work we report results of detailed analyses of experimental 
data on pp and $\rm p \bar p $ elastic scattering cross section 
in terms of their amplitudes for the highest energies available. 
In spite of the simplicity of the form of the profile amplitudes, 
which originates on applications of the stochastic vacuum model,
our analytic representation reproduces the experimental data with 
remarkable precision in the whole t domain,  from the very forward 
scattering  to the structure and position of dips and bumps,  further 
still allowing extension to the tail domain where the perturbative 
three gluon exchange mechanism is expected to dominate. 

It is found that the energy dependence of the fitted parameters is 
quite regular as function of $\ln{s}$, showing that they can 
carry important information on QCD dynamics and proton structure. 

When compared to the description obtained for lower ISR energy data, 
we still observe smooth connection, although  deflection in some 
interpolated lines may suggest  possible additional changes in 
curvatures occurring  in the energy range 
 $ 100 \GeV \leq \sqrt{s} \leq 500 \GeV$. 
The occurrence of disturbed regularity may happen if some new channels 
in the final state of pp or $\rm p \bar p$  collisions start to open 
in this energy range (for example, emergence of deconfined quark-gluon 
droplets). This phenomenon would starts to increase the inelastic 
and the total cross sections. Thus, through the optical theorem,  elastic 
cross section would also be affected.  Therefore, emergence of such a 
critical behavior should reflect in the change in curvature of parameters 
as function of energy. Unfortunately, there exist no data points for this 
energy domain. We expect that pp experimental data in the RHIC energies 
could have a decisive role for the understanding the QCD dynamics at the 
critical region.

The identification of the imaginary and real amplitudes here presented 
may provide a framework for development and control of ingredients of 
microscopic theoretical models that study the underlying dynamics.

\begin{acknowledgments}

The authors wish to thank CNPq, PRONEX and FAPERJ for financial support.  
 
\end{acknowledgments}

\end{document}